\documentclass[twocolumn,showpacs,amsmath,amssymb,pre]{revtex4}

\usepackage{graphicx}
\usepackage{dcolumn}
\usepackage{bm}

\begin{document}

\title{Intra-molecular coupling as a mechanism  for a
liquid-liquid phase transition}  

\author{Giancarlo Franzese$^{1,2}$} 
\altaffiliation[Present address:~SMC-INFM, Dipartimento di Fisica,
Universit\`a ``La Sapienza'', P.le A. Moro 2, I-00185 Roma, Italy]{} 
\author{Manuel I.  Marqu\'es$^1$}
\author{H. Eugene Stanley$^1$} 

\affiliation{%
$^1$~Center for Polymer Studies and Department of Physics,
Boston University, Boston, MA 02215, USA\\
$^2$~Dipartimento di Ingegneria
dell'Informazione, 
Seconda Universit\`a di Napoli, 
INFM UdR Napoli and CG SUN,
via Roma 29 I-81031, Aversa (CE), Italy 
}%

\date{\today}

\begin{abstract} 
We study a model for water with a tunable intra-molecular interaction
$J_\sigma$, using mean field theory and off-lattice Monte Carlo simulations. For all
$J_\sigma\geq 0$, the model displays a temperature of maximum density.
For a finite intra-molecular interaction $J_\sigma > 0$,   
our calculations support the
presence of a liquid-liquid phase transition with a possible liquid-liquid
critical point for water, likely pre-empted by inevitable freezing.
For $J=0$ the liquid-liquid critical point disappears at $T=0$.

\end{abstract}


\maketitle

\section{Introduction}

Water has an anomalous density decrease for isobaric cooling below a
temperature of maximum density (TMD) \cite{Debenedettibook}.  Other
thermodynamic anomalies---such as the rapid increase of the response
functions---can be fit by power laws with apparent singularity well below
the freezing temperature \cite{Debenedettibook}. 
Several interpretations for this
behavior have been proposed, but it is unclear which describes water or if
any describes other anomalous liquids including, among the others, 
S, Se, Te, Cs, Si, Ge, I, C, P, 
SiO$_2$, BeF$_2$
\cite{Volga,russians,Angell_JNCS,Angell,Angell_excitations,Angell_preprint,%
Poole94,Debene,Mishima,Soper,Glosli,katayama,morishita01,Ivan,fmsbs,Lee,Guisoni01}.  

One of the interpretations, the {\em stability-limit
conjecture} \cite{Speedy82_Dantonio87_Debenedetti88}, assumes that the
limits of stability of the superheated, supercooled and stretched liquid
form a single retracing spinodal line in the pressure--temperature
($P$--$T$) plane. This scenario predicts 
a divergence of the response functions at the
supercooled liquid-to-liquid spinodal \cite{Sastry93}.  

The {\em
singularity-free} interpretation \cite{StanleyTeixeira80,Sastry96}
envisages the experimental data represent apparent singularities, due to
anticorrelated fluctuations of volume and entropy.  In this scenario, 
these fluctuations are responsible for the TMD line.

The {\em liquid-liquid
phase transition} hypothesis \cite{Poole} proposes the presence of a
first order line of phase transitions separating two liquid phases
differing in density, the high density liquid (HDL) and the low
density liquid (LDL). In this scenario the HDL-LDL phase transition, possibly
ending in a liquid-liquid critical point, is responsible for the
anomalies.

Although Refs.~\cite{Poole94,Debene}, by tuning parameters of the
corresponding models, predict smooth transitions from the different
scenarios, to help elucidate which is the most reasonable description for
water, we consider a model fluid with {\it inter-molecular} and 
{\it intra-molecular} interactions.
This model, in the particular case of 
a zero intra-molecular interaction, recovers the model
introduced by Sastry et al. ~\cite{Sastry96} that predicts the
singularity-free scenario. Our aim is to understand how the presence
of an intra-molecular interaction changes this prediction.

We perform analytic calculations 
in a mean field approximation and an off-lattice Monte
Carlo (MC) simulation. Our results 
show that a non-zero intra-molecular interaction
gives rise to a HDL-LDL phase transition, with a possible
critical point \cite{Poole}, with the
liquid-liquid critical
temperature decreasing to zero and vanishing 
with the intra-molecular interaction.
Therefore, at least for this model, the singularity-free scenario is obtained
only in the particular case of a zero intra-molecular interaction, while for 
a finite intra-molecular interaction the HDL-LDL phase transition is predicted.
General considerations suggest that the liquid-liquid phase transition for water 
could occur below the glass temperature, i.e. outside the accessible 
experimental range.

The paper is organized as follows: in Sec.~II we define the model
defined on a lattice.
In Sec.~III we describe the equation of state approach in the mean
field approximation and we  present the mean
field results. In Sec.~IV we introduce the off-lattice model, we
describe the MC 
approach and show the simulation results. In Sec.~V we discuss our
results and we give the conclusions.

\section{The lattice model}

The fluid is represented by partitioning the system into $N$ cells of
equal size.  A variable $n_i$ is associated with each cell $i=1,\dots, N$,
with $n_i=1$ if the cell is occupied by a molecule, $n_i=0$ otherwise. 

The inter-molecular interaction~\cite{Sastry96},
\begin{equation} 
{\cal H}\equiv -\epsilon\sum_{\langle i,j\rangle}{n_in_j}
     -J\sum_{\langle i,j\rangle}{n_in_j\delta_{\sigma_{ij},\sigma_{ji}}}
~, 
\label{LG} 
\end{equation} 
has a first term describing the van der Waals attraction
between molecules, where $\epsilon>0$ is the energy gain for
two nearest neighbor (nn) occupied cells and the sum is over all the
possible nn cells.  

The second term in Eq.~(\ref{LG}) accounts for the
dynamic network of hydrogen bonds (HBs) formed by liquid water, with each
molecule typically bonded to four other molecules at low $T$
\cite{Debenedettibook}, with an energy-gain $J>0$ per HB. We consider
cells with size of a water molecule and with four {\em arms}, one per
possible HB.  For the molecule in the cell $i$, the orientation of the arm
facing the cell $j$ is represented by a Potts variable $\sigma_{ij}=1, \dots,
q$, with a finite number $q$ of possible orientations.  Two molecules in
nn cells form a HB only if they are correctly oriented
\cite{Debenedettibook}, i.e., by assumption \cite{Sastry96}, if
$\delta_{\sigma_{ij},\sigma_{ji}}=1$ ($\delta_{a,b}=1$ if $a=b$ and
$\delta_{a,b}=0$ otherwise). 
 
The experimental oxygen-oxygen correlation function shows that a HB is
formed if and only if the inter-molecular distance is within a
characteristic range \cite{Debenedettibook}.  Hence, we assume that 
the formation of a HB
leads to a local volume expansion \cite{Sastry96},
\begin{equation}
V\equiv V_0 + N_{HB} v_{HB} ~,
\label{vol}
\end{equation}
where $V_0$ is the volume of the liquid with no HBs, 
\begin{equation}
N_{HB}\equiv
\sum_{\langle i,j\rangle}{n_in_j\delta_{\sigma_{ij},\sigma_{ji}}} 
\label{Nhb}
\end{equation}
is the
total number of HBs in the system, and $v_{HB}$ is the specific volume per
HB. 

Experiments show that the relative orientations of the arms of a water
molecule are correlated, with the average H-O-H angle equal to $104.45^{o}$ in
an isolated molecule, $104.474^{o}$ in the gas and $106^{o}$ in the high-$T$
liquid \cite{Kern_Hasted_Ichikawa}, suggesting an {\em intra-molecular}
interaction between the arms.  This interaction must be finite, because
the angle changes with $T$, consistent with ab-initio calculations
\cite{Silvestrelli} and molecular dynamics simulations 
\cite{Netz}. Hence, we
introduce the intra-molecular term \cite{fys,fs}
\begin{equation}
{\cal H}_{IM}\equiv
     -{J_\sigma}\sum_i n_i\sum_{(k,l)_i}{\delta_{\sigma_{ik},\sigma_{il}}} ~,
\label{new}
\end{equation}
where for each of the $^{4}C_{2}=6$ different pairs $(k,l)_i$ of the arms
of a molecule $i$, with the appropriate orientation
($\delta_{\sigma_{ik},\sigma_{il}}=1$), there is an energy gain
$J_\sigma>0$. 

For $J_\sigma=0$ we recover the model of Ref.~\cite{Sastry96} that
predicts the singularity-free scenario, and where the
HBs are uncorrelated, inhibiting the orientational long-range order.  We
study the general case with finite $J_\sigma$, by using ($i$) a mean field
approximation and ($ii$) MC simulations. 

\section{The equation of state approach}

The equation of state of our system is implicitly given by
\begin{equation}
U-TS+PV=\mu\sum_i n_i~, 
\end{equation}
where 
\begin{equation}
U\equiv {\cal H}+{\cal H}_{IM}
\label{u}
\end{equation}
is the
total internal energy and $\mu$ is the chemical potential.  From
Eqs.~(\ref{LG})-(\ref{new}), we rewrite the equation of state as
\begin{equation}
TS-PV_0=
-\sum_{\langle i,j \rangle} \epsilon'_{ij}(P,\sigma) n_in_j
-\sum_i\mu_i'(\sigma)n_i ~.
\label{eos}
\end{equation}
Here 
\begin{equation}
\epsilon_{ij}'(P,\sigma)\equiv 
\epsilon+J'(P)\delta_{\sigma_{ij},\sigma_{ji}} 
\end{equation} 
is the effective attraction energy, depending on $P$ and the local arm 
configuration,
\begin{equation}
J'(P)\equiv J-Pv_{HB} 
\end{equation} 
is the effective HB interaction energy due to the
HB volume-increase, and 
\begin{equation}
\mu_i'(\sigma)\equiv 
\mu+J_\sigma\sum_{(k,l)_i}\delta_{\sigma_{ki},\sigma_{li}} 
\end{equation} 
is the effective local chemical potential depending on the local arm
configuration. 

\subsection{The mean field approximation}

The mean field approximation consists in assuming a linear relation
between the number density of liquid cells, 
\begin{equation}
n\equiv \frac{\sum_i n_i}{N}~, 
\end{equation}
and
the density order parameter $m\in[-1,1]$, and between the number density
of arms in the appropriate state for a HB, 
\begin{equation}
n_\sigma \equiv \sum_{\langle i,j \rangle} \delta_{\sigma_{ij},1}~, 
\end{equation}
and the orientational order
parameter $m_\sigma\in[0,1]$, i.e. 
\begin{equation}
n=\frac{1+m}{2} ~, ~~~ n_\sigma=\frac{1+(q-1)m_\sigma}{q} ~.
\label{mf}
\end{equation}

Hence, the molar density $\rho\equiv nN/V$ is
\begin{equation}
\rho=\frac{1+m}{2v_0+4v_{HB}p_{HB}} ~.
\label{rho} 
\end{equation} 
Here $v_0\equiv V_0/N$, and $p_{HB}\equiv n^2p_\sigma$ is the probability
of forming a HB between two nn molecules, where $n^2$ is the probability
of finding two nn molecules and
\begin{equation}
p_\sigma\equiv
n_\sigma^2+(q-1)\left(\frac{1-n_\sigma}{q-1}\right)^2
= \frac{1+(q-1)m_\sigma^2}{q}
\label{psig}
\end{equation}
is the probability of having the facing arms of the two molecules in the
appropriate orientational state for a HB.  

For $T\rightarrow\infty$ we
expect $m_\sigma\rightarrow 0$, hence $p_\sigma\rightarrow 1/q$. For
$T\rightarrow 0$ the finite values of $\epsilon$, $J$ and $J_\sigma$ allow
us to assume a {\it cooperative effect} and an orientational long-range order
in a preferred state, with $m_\sigma\rightarrow 1$, hence
$p_\sigma\rightarrow 1$. 

\subsection{The cooperative effect}

To include the cooperative effect, we consider that each arm
$\sigma_{ij}$ interacts with a mean field $h$ generated by the other three
arms on the same molecule, in addition to the effective interaction
$J'(P)$ with the facing arm $\sigma_{ji}$ on a nn molecule.  Since the
energy is minimized when the arms are in the same orientational state, the
system breaks the symmetry ordering in the preferred Potts state. Hence,
we choose $h$ proportional to $n_\sigma$, to $J_\sigma$ and to the number
of arms generating $h$, i.e.  
\begin{equation}
h\equiv 3J_\sigma n_\sigma~.
\end{equation}
Our results
do not depend on this choice for $h$, and are recovered using higher order
approximations for $h$. 

We assume 
\begin{equation}
p_\sigma=\langle \delta_{\sigma_{ij},\sigma_{ji}} \rangle_h ~,
\label{condition}
\end{equation}
i. e. 
that $p_\sigma$ [Eq.(\ref{psig})], for two nn molecules interacting with the
surrounding, coincides with the probability 
$\langle \delta_{\sigma_{ij},\sigma_{ji}} \rangle_h$,
for the facing arms ($\sigma_{ij}, \sigma_{ji}$) of
two {\em isolated} nn molecules, of being  
in the same orientational state under the
action of the field $h$. By definition is
\begin{eqnarray}
\langle \delta_{\sigma_{ij},\sigma_{ji}} \rangle_h  & \equiv &
\frac{1}{{\cal Z}_h}\sum_{\sigma_{ij},\sigma_{ji}}\delta_{\sigma_{ij},\sigma_{ji}}
\exp\{[J'(P)\delta_{\sigma_{ij},\sigma_{ji}} \nonumber\\
& + & h(\delta_{\sigma_{ij},1}+\delta_{\sigma_{ji},1})]/(k_BT)\} \\
& = &  
\left[
1+(q-1)
\frac{2w_{m_\sigma}+q-2}
{w(w_{m_\sigma}^2+q-1)}
\right]^{-1}  ~ , \nonumber
\label{Psigma}
\end{eqnarray}
where the right-most-side is the explicit calculation of the left-side
with partition function ${\cal Z}_h$. Here the sum is over all the
configurations of the two variables 
$\sigma_{ij}$, $\sigma_{ji}$, and the symbols are
\begin{eqnarray}
w            & \equiv & \exp\left[\frac{J'(P)}{k_BT}\right] ~,\nonumber\\
w_{m_\sigma} & \equiv & 
\exp\left\{\frac{3J_\sigma[1+m_\sigma(q-1)]}{qk_BT}\right\}~,\\
{\cal Z}_h & \equiv & \sum_{\sigma_{ij},\sigma_{ji}}
\exp\left[\frac{J'(P)\delta_{\sigma_{ij},\sigma_{ji}}+ 
h(\delta_{\sigma_{ij},1}+\delta_{\sigma_{ji},1})}{k_BT}\right] ~,\nonumber
\end{eqnarray}
with the Boltzmann constant $k_B$ chosen as unitary hereafter.
As expected for $p_\sigma$, also
$\langle \delta_{\sigma_{ij},\sigma_{ji}} \rangle_h \rightarrow 1/q$ 
for $T\rightarrow \infty$
and $\langle \delta_{\sigma_{ij},\sigma_{ji}} \rangle_h \rightarrow 1$ 
for $T\rightarrow 0$.

Numerically we find that the solution of the Eq.(\ref{condition})
is $m_\sigma^*(T,P)=0$ for $P>P_{max}(T)$, 
and $m_\sigma^*(T,P)>0$ for $P\leq P_{max}(T)$, 
where $m_\sigma^*(T,P)=0$ corresponds to lack of 
orientational order and $m_\sigma^*(T,P)>0$
corresponds to the orientational long-range order.
In this mean field approximation, 
$P_{max}(T)$ turns out to be well described by a decreasing linear
function of $T$.

\subsection{The Gibbs free energy}

Next, we write a mean field expression for the molar Gibbs
free energy 
\begin{equation}
g\equiv u-Ts+Pv = \mu
\end{equation}
 as a function of the two order
parameters $m$ and 
$m_{\sigma}^*$, where
\begin{equation}
u\equiv \frac{U}{nN}=-2[\epsilon n+(Jn+3 J_\sigma)p_\sigma]
\end{equation}
is the
molar energy as derived by 
Eqs.~(\ref{LG})-(\ref{new}) and Eq.~(\ref{u}),
with the mean field approximations 
\begin{eqnarray}
\sum_{\langle i,j\rangle}{n_in_j} & = & n^2~,\nonumber\\
\sum_{\langle i,j\rangle}{n_in_j\delta_{\sigma_{ij},\sigma_{ji}}} & = &
n^2p_\sigma~, \\
\sum_i n_i\sum_{(k,l)_i}{\delta_{\sigma_{ik},\sigma_{il}}} & = & 
np_\sigma~;\nonumber 
\end{eqnarray}
$
v\equiv 1/\rho
$ 
is the molar volume derived by Eq.~(\ref{rho});
\begin{equation}
s\equiv \frac{S_{W}+S_\sigma}{nN}
\end{equation}
is the molar entropy,
\begin{eqnarray}
-\frac{S_W}{k_B N} & = & n\ln n + (1-n) \ln (1-n) \\
-\frac{S_\sigma}{4 k_B n N} & = & n_\sigma^*\ln n_\sigma^* +
(q-1) \left(\frac{1-n_\sigma^*}{q-1}\right) 
\ln
\left(\frac{1-n_\sigma^*}{q-1}\right) 
\nonumber
\end{eqnarray}
are the standard mean
field expressions for the entropy of
$N$ variables $n_i$ and
for the entropy of $4nN$ 
$q$-states Potts variables for the arms, respectively, and
$n_\sigma^*$ is the number density $n_\sigma$ of HBonded arms 
[Eq.~(\ref{mf})] evaluated in 
$m_\sigma^*(T,P)$. 

\subsection{The mean field results}

\begin{figure}
\includegraphics[width=8.5cm,height=7cm,angle=0]{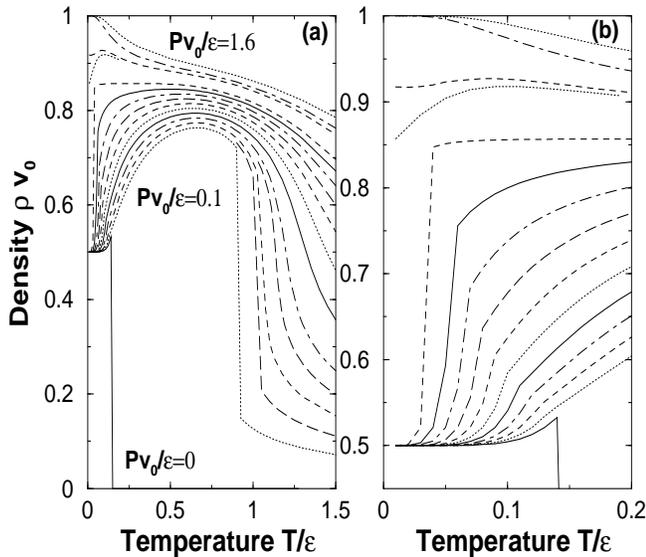}
\caption{
The mean field isobaric molar density $\rho$ as function of
$T$ for (top to bottom) $Pv_0/\epsilon=1.6$, 
1.4, 
1.3, 1.275, 
1.0, 
0.9, 0.8, 0.7, 0.6, 0.5, 0.4,
0.3, 0.25, 0.2, 0.15, 0.1, 0,
for the model with parameters
$q=6$,
$J/\epsilon=0.5$,
$J_\sigma/\epsilon=0.05$,
and
$v_{HB}/v_0=0.5$.
(a) For $0.1\lesssim Pv_0/\epsilon \lesssim 0.25$,
by decreasing $T$, $\rho$ has a discontinuity at high $T$, from a
low value (in the gas phase) to a high value (in the liquid phase), then
$\rho$ has a maximum followed by a  smooth saturation to the finite value
$\rho_{HB}=0.5/v_0$ corresponding to the full-HBonded liquid.
For $0.3 \lesssim Pv_0/\epsilon \lesssim 0.4$, by decreasing $T$, there is no
discontinuity in $\rho$, but there is a maximum in $\rho$ and the
saturation to $\rho_{HB}$. 
For $0.5 \lesssim Pv_0/\epsilon \lesssim 1.2$, by decreasing $T$, $\rho$ has
a maximum and then a discontinuity to $\rho_{HB}$. 
For $1.25 \lesssim Pv_0/\epsilon \lesssim 1.6$, $\rho$ has only a maximum,
and, for higher $P$, $\rho$ regularly increases, by decreasing $T$.
(b) Blowup of the low-$T$ region.
Both discontinuities reported show a first order phase transition, 
each ending in a critical point (Fig.~2).
}
\label{rhot}
\end{figure}

By numerically minimizing $g(T,P)$ with respect to $m$ and
$m_\sigma^*$ with the constraint that $m_\sigma^*$ is
solution of 
Eq.~(\ref{condition}),
we find the equilibrium values of $m(T,P)$ and
$m_\sigma^*(T,P)$. By using Eq.~(\ref{rho}), we find 
of $\rho(T,P)$ at equilibrium (Fig.~\ref{rhot}).

\begin{figure}
\includegraphics[width=8.5cm,height=7cm,angle=0]{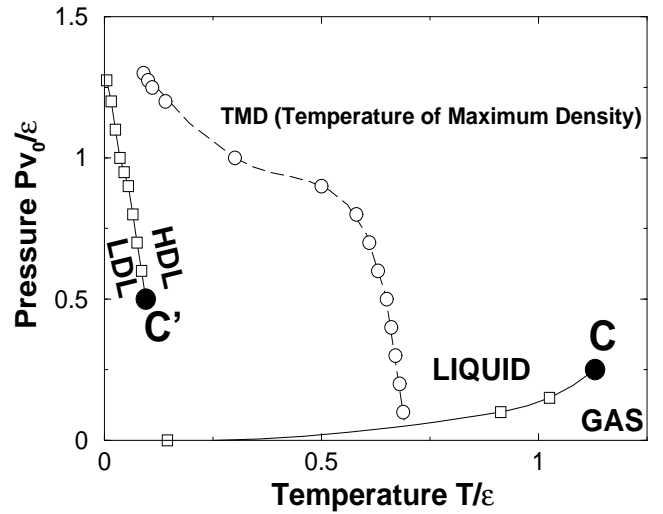}
\caption{
The $P$--$T$ phase diagram for the model with the parameters in Fig.~1.
The squares and the circles are estimated from the discontinuities and
the maxima in Fig.~1, respectively.
Since the symmetry between the two phases separated by a critical
point must be preserved, the high-$T$ discontinuity shows a
gas-liquid phase transition, while the low-$T$ discontinuity shows the
HDL-LDL phase transition. 
The lines are guides for the eyes.
The liquid-gas phase transition line ends in the critical point $C$.
The HDL-LDL phase transition line, with
negative slope, ends in the critical point $C'$.
The dashed line corresponds to the TMD line. 
}
\label{pt}
\end{figure}

At high $P$ the mean field theory predicts that $\rho(T)$ increases when
$T$ decreases (Fig.~1).  At low $P$, for decreasing $T$, the theory
predicts ($i$) a discontinuity in $\rho(T)$, corresponding to the
liquid-gas first order phase transition ending in the liquid-gas critical
point $C$ (Fig.~\ref{pt}), ($ii$) the TMD decreasing for increasing $P$,
($iii$) a discontinuity in
$\rho(T)$ at low $T$, disappearing at lower $P$.  

The first two
predictions are consistent with either the singularity-free scenario or
the liquid-liquid phase transition hypothesis, while the third prediction
is consistent only with the HDL-LDL
first order phase transition hypothesized in the latter scenario. In
particular, the smooth disappearing of the discontinuity at lower $P$ is
consistent with a phase transition line, with a negative slope in the
$P$--$T$ phase diagram, ending in a HDL-LDL 
critical point $C'$ (Fig.~\ref{pt}). 

\section{The off-lattice model}

To show that our mean field predictions are robust, we now use a
completely different approach based on an off-lattice (OL) model
representing a system with a homogeneous
distribution of molecules in the available volume $V$, that we divide in $N$
equivalent cells of volume $V/N$. 

As a consequence of the homogeneity of the system,
for each cell the $N$ degrees of occupancy freedom $n_i$ are set to
$n_i=1$. In analogy with Eq.~(\ref{vol}), the total volume $V$ is
defined as  
\begin{equation}
V\equiv V_0^{OL}+N_{HB}v_{HB}~.
\label{v_ol}
\end{equation}
Here $N_{HB}$ and $v_{HB}$ are defined as in
Eqs.~(\ref{vol})-(\ref{Nhb}), 
but, differently from the lattice case, the volume $V_0^{OL}$,
associated with the total volume of the cells without HBs, is a
continuous variable with the constrains $V_0^{OL}\geq N v_0$, 
where $v_0$ is the hard-core volume of a molecule.

Also,
following the lattice model, the molecules have four arms described by
four $q$-state variables $\sigma_{ij}$, with the HB interaction defined by
the second term in Eq.~(\ref{LG}) and the intra-molecular interaction by
Eq.~(\ref{new}). These interactions are both independent of the distance
among first-neighbor molecules and depend only on the arms orientation
$\sigma_{ij}$.  

In this off-lattice model, the average distance between two
molecules $r$ is a continuous variable, so we replace the first term in
Eq.~(\ref{LG})  with
\begin{equation}
U_W(r)\equiv \left\{
\begin{array}{lll}
\infty & \mbox{  for  } & r \leq R_0\\
\epsilon \left[\left(\frac{R_0}{r}\right)^{12}-\left(\frac{R_0}{r}\right)^6\right] & \mbox{  for  } & r>R_0 ~,
\end{array}\right.
\label{Uw}
\end{equation}
where $R_0\equiv \sqrt{v_0}$ is the hard-core diameter 
of each molecule. In analogy
with the Eq.~(\ref{LG}), we consider this off-lattice van der Waals energy
independent of the HB expansion, so  in
$U_W(r)$ we use 
\begin{equation}
r\equiv \sqrt{V_0^{OL}/N} ~.
\end{equation}

\subsection{The Monte Carlo simulation}

\begin{figure}
\includegraphics[width=8.5cm,height=7cm,angle=0]{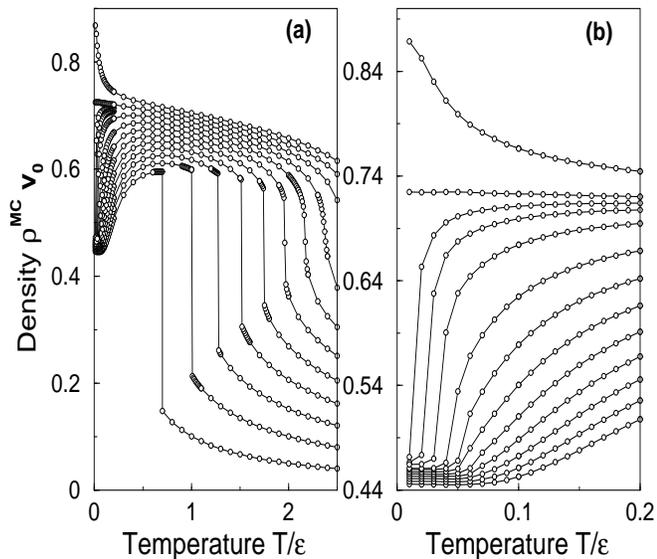}
\caption{MC isobaric density $\rho^{MC}(T)$  
for $N=10^4$ molecules, for the off-lattice model 
with parameters as in Fig.~1. 
We show only the isobars for (bottom to top)
$Pv_0/\epsilon=0.1$, 0.2, 0.3, 0.4, 0.5, 0.6, 0.7, 0.8, 0.9, 0.95, 0.975, 1,
1.1.  
(a) The qualitative behavior is as described in Fig.~1.
(b) Blowup of the low-$T$ region.}
\label{mc1}
\end{figure}

We perform MC simulations, in two dimensions \cite{noteMC}, at
constant $N$, $P$, $T$ and variable $V$ ($NPT$ ensemble) with $N\in
[10^2,10^4]$.  
The MC dynamics consists in updating the variables $\sigma_{ij}$
and the variable $V_0^{OL}$, accepting the new state with probability
$\exp[-(\Delta U_W + P\Delta V/k_BT)]$ if $\Delta U+P\Delta V>0$, or with 
probability 1 if $\Delta U + P\Delta V<0$. Here
$\Delta U\equiv \Delta (U_W+{\cal H}_{IM})$ and $\Delta
V$ are the changes of total internal energy and total volume
Eq.(\ref{v_ol}), respectively, after the update.
Our results for the average density $\rho^{MC}\equiv N/V$,
averaged on $6\times 10^5$ MC steps
after $1.2\times 10^5$ MC steps of thermalization at each $T$, are
qualitatively consistent with the mean field prediction (Fig.~\ref{mc1}). 

\begin{figure}
\includegraphics[width=8.5cm,height=7cm,angle=0]{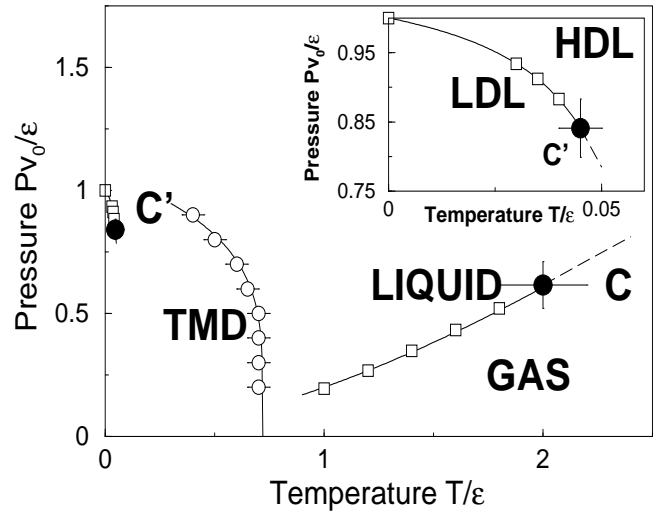}
\caption{$P$--$T$ phase diagram calculated by MC simulations, 
for the off-lattice model as in Fig.~3.
Squares are the 
$N\rightarrow \infty$ estimates for the points on the coexisting lines.
Full circles are 
the critical points $C$ and $C'$.
Points on the TMD line (open circles) are estimated from 
the $N\rightarrow \infty$ extrapolation of the
maxima of $\rho^{MC}(T,N)$.
Dashed lines indicate the position of $K_T^{max}$ emanating from $C$
and $C'$. Full lines are guides
for the eyes.  
Where not shown, errors are smaller then the symbol size.
Inset: blowup of the HDL-LDL phase transition region.
The full line is given by the empiric expression
$P_{max}=P_0-[a/(T_0-T)]$ where $P_0v_0/\epsilon=1.087$,
$av_0/\epsilon^2=0.006$ and $T_0/\epsilon=0.07$.
}
\label{mc2}
\end{figure}

By MC simulations, we find ($i$) for $J_\sigma=0$, no liquid-liquid phase
transition and the TMD line, consistent with mean field in
Ref.~\cite{Sastry96};  
($ii$) for $J_\sigma \gg J$, for any $P$ in the liquid phase, at $T$ below
the TMD line, a discontinuity in density, suggesting a first order phase
transition along a line with a negative slope in the $P$--$T$ phase
diagram, consistent with mean field for $J_\sigma\rightarrow \infty$
\cite{fys,fs};  ($iii$) for $0<J_\sigma<J$, a phase transition line ending in a
critical point $C'$ (Fig.~\ref{mc2}), 
and occurring at increasing $P$ and decreasing $T$ for decreasing
$J_\sigma$.
 
To verify that the jumps found in the MC density are marking a first order
phase transition, instead of a narrow continuous phase transition, we
study the isothermal compressibility,
\begin{equation}
K_T\equiv -\frac{1}{V}\left(\frac{\partial V}{\partial P}\right)_T~.
\end{equation}
Its maximum $K_T^{max}$, where 
\begin{eqnarray}
(\partial K_T/\partial V)|_T & = & 0 \nonumber\\
(\partial^2 K_T/\partial V^2)|_T & > & 0 ~,
\end{eqnarray}
increases linearly with the number
of particles $N$ at a first order phase transition \cite{challa86},
while at a second order phase transition 
$K_T^{max}$ is proportional to both $N$ and the fluctuation of the
ordering parameter, scaling as a power of $N$ \cite{wilding}.
Therefore, the finite size scaling analysis on $K_T^{max}(N)$ allows
us to discriminate between a continuous and a discontinuous phase
transition.

To estimate the maximum with a great precision, we use a continuous $T$
algorithm, the histogram reweighting method \cite{Ferrenberg_Swendsen}.
By checking the minimum $T$ and the maximum $P$ where the behavior of
$K_T^{max}(N)$ fails to be linear, we 
estimate the critical point at $T_{C'}/\epsilon=0.045\pm0.005$ 
and $P_{C'}v_0/\epsilon=0.841\pm0.042$ for $N\rightarrow \infty$
(inset Fig.4).
 
Next, we obtain the coexisting lines by extrapolating to $N \rightarrow
\infty$ the values $P(T,N)$ corresponding to $K_T^{max}(N)$ at fixed $T$,
both for $T\leq T_C$ and $T\leq T_{C'}$.
Furthermore, our results are consistent with the necessary condition
that the $K_T^{max}(T)$ line emanates from the critical point
\cite{Sastry96,Saika00}, both for $T>T_{C'}$ and for $T>T_C$ (Fig.4). 

\section{Discussion and conclusions}

The mean field and the MC results for our water-like
fluid model for a finite intra-molecular interaction $J_\sigma$
show qualitatively the same phase diagram. In both approaches the TMD
line decreases with increasing $P$, consistent with the experiments
\cite{Angell_JNCS}. In particular, both approaches predict
a first order phase transition in the liquid phase, occurring 
at low $T$ and at a pressure $P_{max}(T)$ decreasing for
increasing $T$.
For $J_\sigma<J$  we find that the first order phase
transition line ends in a critical point, separating, by necessity,
two phases with the same symmetry, in this case two liquids, HDL and LDL,
being the critical point in the liquid phase.
Three considerations are in order here. 

The {\it first consideration} is related to the
comparison with the result of Ref.~\cite{Sastry96},
here recovered for $J_\sigma=0$. 
The Sastry et al. model \cite{Sastry96}, upon HB formation, accounts for 
($i$) inter-molecular orientational correlation [Eq.(\ref{LG})];
($ii$) local expansion with lowering temperature [Eq.(\ref{vol})];
($iii$) anticorrelation between $V$ and $S$, because
the formation of HBs decreases the number of possible
orientational configurations for the system, hence the entropy $S$
decreases for increasing $N_{HB}$, i.e. for increasing $V$. 
This is expected in a system with a density anomaly, 
because $(\partial V/\partial T)_P<0$ implies $(\partial S/\partial
V)_T<0$.  
Finally, the Sastry et al. model assumes the arms of a molecule
completely independent ($J_\sigma=0$), and predicts the singularity-free
scenario.

We tested, by preliminary MC calculations, that for
$J_\sigma\rightarrow 0$ the HDL-LDL critical point $C'$ moves to lower
$T$ and to $P_{max}(T=0)$. In particular, the phase transition
disappears at $T=0$ for $J_\sigma=0$, while the effect
of the decreasing $J_\sigma$ on the location of the TMD line is weak.
Although these results require a longer analysis, beyond the scope of
the present work, they show that the predictions of  
Ref.~\cite{Sastry96} are recovered in the
limit $J_\sigma=0$, confirming the validity of our MC approach \cite{note}. 
We, therefore, conclude that in this model the presence of a finite
$J_\sigma$ is 
responsible for the appearance of the first order phase transition,
with a possible HDL-LDL critical point $C'$.

The {\it second consideration} is related to the possibility that this
low-$T$ HDL-LDL phase transition is pre-empted by inevitable freezing 
in real
water. Recent analysis of the realistic model for water ST2
\cite{Ivan} suggests that the HDL-LDL critical point may occur
above the glass temperature $T_g$, though as yet still
outside the easily accessible experimental range.

\begin{table*}  
\caption{Characteristic temperatures and pressures for real water and for the 
present model, for the gas-liquid critical point ($T_C,P_C$), the TMD at ambient
pressure ($T^*,P^*$) and the glass temperature $T_g$ at ambient 
pressure. The ratios $P^*/P_C$ and $T_g/T_C$ are not available for the model.
We assume that the corresponding H$_2$O values are valid also for the model, and we 
use these values to estimate $P^*$ and $T_g$, respectively, for the model.
Temperatures are measured in K for H$_2$O and in $\epsilon$
for the model. Pressures are measured in MPa for H$_2$O and in
$\epsilon/v_0$ for the model.}    
\begin{ruledtabular}
\begin{tabular}{c | c c c c c c c c}  
       & 
$P_C$      & 
$T_C$      & 
$P^*$            & 
$P^*/P_C$   & 
$T^*$            & 
$T^*/T_C$     & 
$T_g$      & 
$T_g/T_C$ \\ 
\hline
H$_2$O & 
22.064     & 
647.14	    & 
0.10133~\cite{ts} & 
$4.6~10^{-4}$ & 
277~\cite{ts}     & 
0.428         & 
136~\cite{tg} & 
0.21 \\ 
model  & 
$0.6\pm 0.1$ & 
$2.0\pm 0.2$ & 
$\sim 2.76~10^{-4}$     & 
-           & 
$0.7 \pm 0.1$ & 
$0.35 \pm 0.06$ & 
$\sim 0.42$  & 
-
\end{tabular}
\end{ruledtabular}
\label{tab1}
\end{table*}  

We compare our MC results with data for real water.
From the location of the liquid-gas critical point $(T_C,P_C)$ and the
TMD line at ambient pressure $(T^*,P^*)$ (Table~\ref{tab1}), we find the
ratios $P^*/P_C$ and $T^*/T_C$ in real water.
By assuming that the same $P^*/P_C$ holds in our MC case, we calculate
the corresponding $P^*\sim 2.76 ~ 10^{-4}\epsilon/v_0$ 
in our model and then we estimate
the $T^*$ corresponding to $P^*$ from the TMD line in the MC phase
diagram (Fig.~\ref{mc2}). 
In this way we find a ratio $T^*/T_C$ from the MC results that is
consistent with the real water data, suggesting the validity of our
assumption on $P^*/P_C$.

Therefore, we use the same kind of assumption also to estimate the glass
temperature $T_g$ for our phase diagram. In particular,   
from real water data we obtain the ratio $T_g/T_C$ at ambient pressure
and, assuming 
that it holds also for our model, we estimate 
$T_g/\epsilon\sim 0.42$ at $Pv_0/\epsilon\sim 2.76~10^{-4}$ 
for the MC phase diagram.
Hence, for our model with the parameters chosen in this paper, 
is $T_g>T_{C'}$, i.e. the HDL-LDL critical temperature at
$P_{C'}\simeq 0.841\epsilon/v_0$ is below the
glass temperature at $P^*\sim 2.76~10^{-4}\epsilon/v_0$. 
From the study of the phase diagram of real water \cite{Angell_JNCS}
is reasonable to expect that $T_g(P)$ decreases for increasing $P$,
therefore our analysis does not exclude that 
$T_{C'}$ is above $T_g(P_{C'})$. However, by considering
a very large $J_\sigma$, such that the HDL-LDL 
critical pressure 
is $\sim 2.76~10^{-4}\epsilon/v_0$, we can compare $T_{C'}$
and $T_g$ at the same pressure $P^*$.
Our preliminary results show that the 
HDL-LDL critical temperature is in this case very close to $T_g$.

As a consequence of this analysis, our model supports the possibility
that the HDL-LDL critical point is located deep into the
supercooled region, below or close to the glass 
temperature, depending on the value of $J_\sigma$. Therefore, the 
liquid-liquid phase transition could be pre-empted, if 
our model is representative of the thermodynamic properties of the
real system. This results is analogous to what has been proposed for
silica \cite{Ivan}, another liquid with density anomaly,
suggesting that the present model could provide a general theoretical
framework for anomalous molecular liquids.

The {\it third consideration} is about the role of the tetrahedrality
in determining the properties of anomalous liquids. For the present
model we do not consider a tetrahedral geometry in the two-dimensional
MC approach, and the geometry is not explicitly defined in the mean
field approach. Nevertheless, our results are consistent with the
experimentally accessible phase diagram of real water, suggesting that
the tetrahedral network is not an essential feature for the anomalous
behavior of water-like liquids. 

This conclusion is consistent with
what has been observed by Angell in Ref.~\cite{Angell_preprint} and is
well described by a general cooperative model \cite{Angell_excitations} 
with a generic drive to phase separate the excitations into distinct
regions of space (clustering). In our model the drive is given by the
intra-molecular interaction, that mimics the geometrical drive in
tetrahedral liquids even if is not necessarily limited to the
tetrahedral case. 

In conclusion, we studied the effect of an intra-molecular interaction
$J_\sigma$ in a model for anomalous molecular liquids with a mean
field approach, valid for  $J_\sigma>0$, and with an off-lattice MC simulation.
For $J_\sigma>0$ we found a HDL-LDL phase transition while our MC
results confirm that for $J_\sigma=0$
the singularity-free scenario holds \cite{Sastry96}. 
Hence, the two
interpretations originate from the same mechanism with a different
hypothesis on the intra-molecular interaction; the latter is strictly
valid only for $J_\sigma=0$. Within this framework, the most
reasonable scenario for water includes a HDL-LDL phase transition,
probably hindered by inevitable freezing. Our results suggest also
that the tetrahedrality is not essential to understand the anomalous
behavior in water-like liquids.

\begin{acknowledgments}
We thank C.A. Angell, M. Barbosa, A. Scala, F. Sciortino, and M. Yamada 
for valuable discussions and 
NFS for support.
M. I. Marqu\'es thanks the postdoctoral financial support of
the Spanish Ministry of Education. 
\end{acknowledgments}


\begin{thebibliography}{99}  

\bibitem{Debenedettibook} 
P. G. Debenedetti, 
{\it Metastable Liquids: Concepts and Principles}
(Princeton University Press, Princeton, 1996).

\bibitem{Volga} 
For a review see 
Proceedings of the NATO Advances Research Workshop, Volga River, on
{\it New Kinds of Phase Transitions: Transformations in Disordered
Substances}, 
edited by V. V. Brazhkin, S. V. Buldyrev, V. N. Ryzhov, and H. E. Stanley 
(Kluwer Academic Publishers, Dordrecht, 2002).

\bibitem{russians}
V. M. Glazov, S. N. Chizhevskaya , and S. B. Evgen'ev, Russian J. of
Phys. Chem. {\bf 43}, 201 (1969);
F. Spaepen and D. Turnbull, AIP Conf. Proc. {\bf 50}, 73 (1979);
B. G. Bagley and H. S. Chen, {\it ibid.} {\bf 50}, 97 (1979);
I. L. Aptecar, Soviet Phys. Doklady {\bf 24}, 993 (1979).

\bibitem{Angell_JNCS}
C. A. Angell, S. Borick, and M. Grabow,
J. Non-Cry. Sol. 
{\bf 207}, 463 
 (1996).

\bibitem{Angell}
P. H. Poole, M. Hemmati, and C. A. Angell,
Phys. Rev. Lett.
{\bf 79} 2281 
 (1997);
C. A. Angell, R. D. Bressel, M. Hemmati, E. J. Sare, and J. C. Tucker,
Phys. Chem. Chem. Phys.,  
{\bf 2}, 1559 
 (2000).

\bibitem{Angell_excitations}
C. A. Angell,
J. Phys.: Condes. Matter,
{\bf 12} 6463 
 (2000);
C. A. Angell and C. T. Moynihan, Met. Mat. Trans. B, {\bf 31}, 587
(2000);
C. A. Angell and R. J. Rao, J. Chem. Phys. {\bf 57}, 470 (1972).

\bibitem{Angell_preprint}
C. A. Angell, preprint of Ovshinsky 80th birthday celebration paper, in
press (2002).


\bibitem{Poole94} 
P. H. Poole,  
F. Sciortino, T. Grande, H. E. Stanley, and C. A. Angell, 
Phys. Rev. Lett. {\bf 73}, 1632 (1994).


\bibitem{Debene}
S. S. Borick, P. G. Debenedetti, and S. Sastry, 
J. Phys. Chem. {\bf 99}, 3781 (1995);
%
C. J. Roberts, A. Z.  Panagiotopulos, and P. G. Debenedetti, 
Phys. Rev. Lett. {\bf 77}, 4386 (1996);
 %
C. J. Roberts and P. G. Debenedetti,  
J. Chem. Phys. {\bf 105}, 658 (1996);
 %
T. M. Truskett, P. G. Debenedetti, S. Sastry, and S. Torquato,
{\em ibid.}
{\bf 111}, 2647 (1999).

\bibitem{Mishima}
O. Mishima,
Phys. Rev. Lett. 
{\bf 85}, 334
(2000);
O. Mishima and Y. Suzuki,
Nature (London) {\bf 419}, 599 (2002).

 %
\bibitem{Soper}
A. K. Soper and M. A. Ricci, 
Phys. Rev. Lett. 
{\bf 84}, 2881 (2000);
J. L. Finney, A. Hallbrucker, I. Kohl, A. K. Soper, D. T. Bowron,
{\em ibid.} {\bf 88}, 225503 (2002).

\bibitem{Glosli}
J. N. Glosli and F. H. Ree, 
Phys. Rev. Lett. {\bf 82}, 4659 (1999);

\bibitem{katayama}
Y. Katayama, T. Mizutani, W. Utsumi, O. Shimomura, M. Yamakata,
and K. Funakoshi, 
Nature (London) {\bf 403}, 170 (2000);
Y. Katayama, K. Tsuji, H. Kanda, H. Nosaka, K. Yaoita, T. Kikegawa,
O. Shimomura, J. Non-Cry. Sol. {\bf 207}, 451 (1996).
 %

\bibitem{morishita01}
T. Morishita,
Phys. Rev. Lett. 
{\bf 87}, 105701
(2001).

\bibitem{Ivan}
I. Saika-Voivod, F. Sciortino, and P. H. Poole,
Phys. Rev. E {\bf 63}, 011202 (2000).

\bibitem{fmsbs}
G. Franzese,  G. Malescio, A. Skibinsky, S. V. Buldyrev, and
H. E. Stanley,
Nature (London)
{\bf 409}, 692 (2001).
 
\bibitem{Lee}
H. K. Lee and R. H. Swendsen, Phys. Rev. B {\bf 64}, 214102 (2001).

\bibitem{Guisoni01}
N. Guisoni and V. B. Henriques, 
J. Chem. Phys. {\bf 115}, 5238
(2001).


\bibitem{Speedy82_Dantonio87_Debenedetti88}
R. J. Speedy, J. Phys. Chem. {\bf 86}, 3002 (1982);
M. C. D'Antonio and P. G. Debenedetti, J. Chem. Phys. {\bf 86}, 2229
(1987).

\bibitem{Sastry93}
S. Sastry, F. Sciortino, and H. E. Stanley, J. Chem. Phys. {\bf 98}, 9863 (1993).

\bibitem{StanleyTeixeira80} 
H. E. Stanley and J. Teixeira, 
J. Chem. Phys. {\bf 73}, 3404 (1980).

\bibitem{Sastry96} 
S. Sastry, P.G.  Debenedetti, F. Sciortino,  and H. E. Stanley, 
Phys. Rev. E {\bf 53}, 6144 (1996);
L. P. N. Rebelo, P. G. Debenedetti, and S. Sastry, 
J. Chem. Phys. {\bf 109}, 626 (1998);
E. La Nave, S. Sastry, F. Sciortino, and P. Tartaglia, 
Phys. Rev. E {\bf 59}, 6348 (1999).

\bibitem{Poole} 
P. H. Poole, F. Sciortino, U. Essman,  and H. E. Stanley,
Nature (London) {\bf 360}, 324 (1992).

\bibitem{Kern_Hasted_Ichikawa} 
C. W. Kern and M. Karplus, {\it Water A comprehensive treatise, Vol. 1} 
(Plenum Press, New York, 1972) F. Franks ed., pp 21-91;
J. B. Hasted, {\em ibid.},
pp 255-309;
K. Ichikawa, Y. Kameda, T. Yamaguchi, H. Wakita, and M. Misawa,
Mol. Phys. {\bf 73}, 79 
(1991). 

\bibitem{Silvestrelli}
P. L. Silvestrelli and M. Parrinello, J. Chem. Phys. {\bf 111}, 3572
(1999).

\bibitem{Netz}
P. A. Netz, F. Starr, M. C. Barbosa, and H. E. Stanley, 
cond-mat/0201138 preprint (2002). 

\bibitem{fys}
G. Franzese, M. Yamada, and H. E. Stanley
in {\it Statistical Physics} (AIP Melville, NY, 2000) 
M. Tokuyama and H. E. Stanley ed., pag. 281-288.

\bibitem{fs}
G. Franzese and H. E. Stanley, 
J. Phys-Condens. Mat. {\bf 14}, 2201 
(2002).

\bibitem{noteMC}
This choice turns out  reasonable, since the
MC results agree with the mean field
predictions, expected to be valid for large dimensions.

\bibitem{challa86}
M. S. S. Challa, D. P.Landau, and K. Binder, Phys. Rev. B {\bf 34},
1841 (1986). 

\bibitem{wilding}
N. B. Wilding and K. Binder, Physica A {\bf 231}, 439 (1996).

\bibitem{Ferrenberg_Swendsen}
A. M. Ferrenberg and R. H. Swendsen, 
Phys. Rev. Lett. {\bf 61}, 2635
(1988).

\bibitem{Saika00}
See note [55] in Ref.~\cite{Ivan}.

\bibitem{note}
Note that our mean field approach is valid only for the case
$J_\sigma>0$, because it {\it assumes } the symmetry breaking for the
Potts variables. For $J_\sigma=0$ the valid mean field approach is the
one presented in Ref.~\cite{Sastry96}, that gives rise to 
the singularity-free interpretation. 

\bibitem{ts}
R. A. Fine and F. J. Millero, J. Chem. Phys. {\bf 59}, 5529 (1973).

\bibitem{tg}
G. P. Joharim A. Hallbrucher, and E. Mayer, Nature (London) {\bf 330},
552 (1987).


\end{thebibliography}
\end{document}